\documentclass[aps,pre,superscriptaddress,amsmath,amssymb,twocolumn]{revtex4}
\usepackage{float}
\usepackage[caption = false]{subfig}
\usepackage{graphicx}
\usepackage{amsmath}
\usepackage{hyperref}
\usepackage{color}
\usepackage{bigints}
\usepackage{mathrsfs}
\usepackage{algorithm}
\usepackage{algpseudocode}
\usepackage{multirow}
\usepackage{tabularx}
\usepackage[normalem]{ulem}
\definecolor{darkblue}{RGB}{8,81,156}
\hypersetup{colorlinks,breaklinks,
        linkcolor=darkblue,urlcolor=darkblue,
       anchorcolor=darkblue,citecolor=darkblue}

\date{\today}

\allowdisplaybreaks

\definecolor{dark-purple}{RGB}{118,42,131}
\definecolor{dark-green}{RGB}{27,120,55}
\definecolor{light-purple}{RGB}{231,212,232}
\definecolor{LIGHT-PURPLE}{RGB}{194,165,207}
\definecolor{light-green}{RGB}{168,216,183}
\definecolor{gray}{RGB}{186,186,186}
\definecolor{super-dark-green}{RGB}{0,69,41}
\definecolor{super-dark-purple}{RGB}{63,0,125}
\definecolor{super-dark-blue}{RGB}{8,48,107}
\definecolor{super-dark-red}{RGB}{165,0,38}

\definecolor{super-dark-purple}{RGB}{64,0,75}
\definecolor{super-dark-green}{RGB}{0,68,27}

\usepackage{tabularx,booktabs} 

\usepackage{array}
\newcolumntype{L}[1]{>{\raggedright\let\newline\\\arraybackslash\hspace{0pt}}p{#1}}
\newcolumntype{C}[1]{>{\centering\let\newline\\\arraybackslash\hspace{0pt}}m{#1}}
\newcolumntype{R}[1]{>{\raggedleft\let\newline\\\arraybackslash\hspace{0pt}}m{#1}}

\usepackage{amsthm}
\usepackage{amstext}
\usepackage{amsmath}
\usepackage{setspace}
\usepackage{esint}
\usepackage{wasysym}
\usepackage{stackrel}
\usepackage{textcomp}
\usepackage{cases}
\usepackage[utf8]{inputenc}
\usepackage[toc,page]{appendix}
 \usepackage[OT2, T1]{fontenc}
 \usepackage[english]{babel}

 \usepackage[]{amsfonts} 
 \usepackage[]{amssymb} \usepackage[]{latexsym}
 \usepackage{graphicx,color} \usepackage{amsthm}
 \usepackage{fancyhdr} 

\theoremstyle{plain}

\theoremstyle{definition}

\theoremstyle{remark}

\usepackage{mathrsfs}

\usepackage{calligra}
\DeclareMathAlphabet{\mathcalligra}{T1}{calligra}{m}{n}

\setcitestyle{super}


\usepackage{mathrsfs}

\usepackage{calligra}
\DeclareMathAlphabet{\mathcalligra}{T1}{calligra}{m}{n}

\begin{document}

\title{Ideal Conductor Model: An analytical finite-size correction for non-equilibrium molecular dynamics simulations of ion transport through nanoporous  membranes}
\author{Brian A. Shoemaker}
\affiliation{Department of Chemical and Environmental Engineering, Yale University, New Haven, CT  06520}
\author{Tiago S. Domingues}
\affiliation{Department of Chemical and Environmental Engineering, Yale University, New Haven, CT  06520}

\author{Amir Haji-Akbari}
\email{amir.hajiakbaribalou@yale.edu}
\affiliation{Department of Chemical and Environmental Engineering, Yale University, New Haven, CT  06520}

\date{\today}

\begin{abstract}
\noindent
Modulating ion transport through nanoporous membranes is critical to many important chemical and biological separation processes. The corresponding transport timescales, however, are often too long to capture accurately using conventional molecular dynamics (\textsc{Md}). Recently, path sampling techniques such as forward-flux sampling (\textsc{Ffs}) have emerged as attractive alternatives for efficiently and accurately estimating arbitrarily long ionic passage times. Here, we use non-equilibrium \textsc{Md} and \textsc{Ffs} to explore how the kinetics and mechanism of pressure-driven chloride transport through a nanoporous graphitic membrane are affected by its lateral dimensions. We not only find  ionic passage times and free energy barriers to decrease dramatically upon increasing the membrane surface area, but also observe an abrupt and discontinuous change in the locus of the transition state. These strong finite size effects arise due to the cumulative effect of the periodic images of the leading ion entering the pore on the distribution of the induced excess charge at the membrane surface in the feed. By assuming that the feed  is an ideal conductor, we analytically derive a finite size correction term that can be computed from the information obtained from a single simulation and successfully use it to obtain corrected  free energy profiles with no dependence on system size. We then estimate ionic passage times in the thermodynamic limit by assuming an Eyring-type dependence of rates on barriers with a size-independent prefactor. This approach constitutes a universal framework for removing finite size artifacts in molecular simulations of ion transport through nanoporous membranes and biological channel proteins.
\end{abstract}

\maketitle

\section{Introduction}

\begin{figure}
	\centering
	\includegraphics[width=.3838\textwidth]{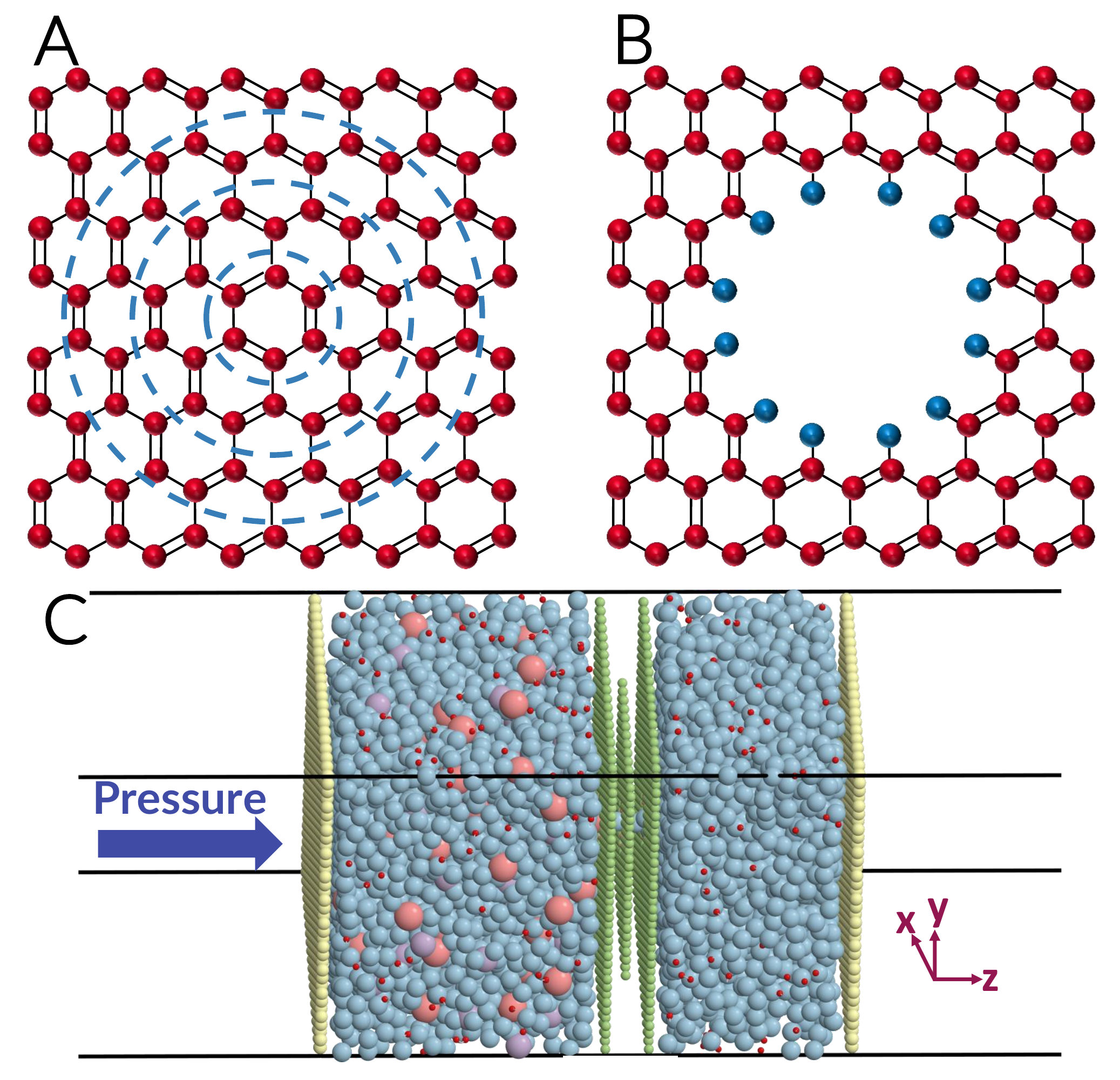}
	\caption{\label{fig:FigCutPassivatedSetup} (A) A bond configuration giving rise to circular pores with single and double bonds drawn onto a graphene sheet and closed loops intersecting the single bonds only. (B) Removing the carbons within the loop interior and passivating the carbons with missing bonds with hydrogens yields the circular pore considered in this work. (C) Schematic representation of a filtration system comprised of a three-layer graphitic membranes with a pore geometry and chemistry given in (B).  Graphene pistons are placed at both ends to apply a pressure gradient of 196~atm. The feed and filtrate are filled with water molecules (light blue), and sodium (light purple) and chloride (dark orange) ions as described in the text.}
	\vspace{-15pt}
\end{figure}

\noindent
Membranes are crucial for maintaining stable steady states within compartmentalized systems by modulating the transport of different molecules and ions.\cite{matsuura2020synthetic} Examples include biological cells and organelles, and industrial  separation processes, such as water desalination and liquid and gas separation. What determines a membrane's efficacy in all such applications is its selectivity,~i.e.,~its ability to preferentially impede the passage of certain chemical entities.\cite{WerberEnvironSciTechnoloLett2016, ZhouSciAdv2020, EpszteinNatNanotechnol2020} How a membrane's molecular structure impacts its selectivity, however, is far from fully understood, mostly due to the inability of existing experimental techniques in the molecular-level probing of solute transport through membranes. In recent years, molecular simulations have emerged as attractive tools for investigating the structure-selectivity relationship at a molecular level.\cite{HeiranianNatComm2015, CohenNanoLett2016, YanNanoscale2017, ZhangPCCP2020, HussainJChemPhys2020, MalmirMatter2020, GuptaNanoscale2020} Simulating membrane separation processes generally requires utilizing techniques such as non-equilibrium molecular dynamics (\textsc{Nemd})\cite{CohenNanoLett2012} as such processes are almost always conducted under non-equilibrium conditions. While \textsc{Nemd} simulations have been extensively used for studying different aspects of driven transport across membranes, they are usually inefficient in accessing transport timescales that exceed $\sim10^{-5}$~s. Probing such timescales has become possible only recently via combining \textsc{Nemd} with advanced sampling techniques. In particular, we recently employed\cite{MalmirMatter2020} jumpy forward flux sampling (j\textsc{Ffs})\cite{HajiAkbariJChemPhys2018} to accurately and efficiently probe arbitrarily long transport timescales for ions traversing a graphitic membrane with sub-nm nanopores.

Despite these advancements, molecular simulations of membrane separation are still at their infancy, and their sensitivity to system setup and other implementation details is yet to be fully investigated. One such overlooked aspect is \emph{finite size effects},~i.e.,~the extent to which transport kinetics and mechanism are affected by the size of the simulation box. Finite size effects are known to systematically impact estimates of thermodynamic\cite{BinderFerroelectrics1987, MonJChemPhys1992, HorbachPhysRevE1996, AguadoJChemPhys2001, OreaJCP2005, MastnyJChemPhys2007, BiscayJCP2009, BurtJPhysChemC2016}, structural~\cite{SalacusePhysRevE1996} and transport~\cite{YehJPCB2004,  BotanMolPhys2015, JamaliJChemTheoryComput2018} properties as well as rates of rare events such as nucleation.\cite{HussainJCP2021, HussainJCP2022} While all such estimates eventually converge to their corresponding value in the thermodynamic limit upon increasing the system size, the rate of convergence tends to be considerably slower for activated phenomena. Conducting a systematic analysis of finite size effects in membrane transport is critical since failing to do so \emph{might} potentially lead to large errors in estimating solvent-solute and solute-solute selectivity, as transport timescales of different chemical species might exhibit vastly different scalings with system size.

In this work, we use our previously developed approach based on j\textsc{Ffs}\cite{MalmirMatter2020} to probe the passage of chloride ions dissolved in an aqueous NaCl solution through a sub-nm cylindrical pore etched within graphitic membranes of identical thicknesses and pore sizes but differing surface areas. We observe that ionic passage times have a strong dependence on membrane cross-sectional area and change by as much as six orders of magnitude within the range of membrane areas considered here. Water flux, on the other hand, is virtually insensitive to system size, which means that finite size effects result in vast overestimation of selectivity and salt rejection rates.  We confirm that such strong finite size effects are artifacts of periodic boundary conditions and are caused by the accumulation at the membrane surface of the surplus charge induced upon removing a chloride from the feed. Assuming that the feed is an ideal conductor, we develop a theory that correctly estimates the systematic error introduced as a result of periodicity and yields corrected generalized free energy profiles that match one another and exhibit no dependence on system size.  Our theoretical model provides a general framework that can be universally utilized to correct for finite size artifacts in simulations of ion transport through nanoporous membranes and biological channel proteins. 

\section{Methods}
\label{section:methods}

\subsection{System Setup and Relaxation}
\label{section:methods:setup}

\noindent 
The membranes considered in this work are comprised of three layers of graphite with sub-nm pores etched within them using the following procedure. In order to create a pore of desired size, carbons are removed as needed from each full graphene sheet and hydrogens are attached to the carbons with missing bonds at a distance of 1.0919~\AA~along the lines connecting them to their now-missing neighbors. However, care must be taken in this approach in order to ensure that every carbon at the pore boundary only has one unpaired valence electron  since a hydrogen can only form a single bond. (This detail has been overlooked in several earlier works,\cite{CohenNanoLett2012, CohenNanoLett2016} including our previous paper.\cite{MalmirMatter2020})  We therefore propose a new method to assure the chemical fidelity of the constructed membranes. First, single and double bonds are drawn on a graphene sheet such that each carbon atom has one double bond and two single bonds (Figure ~\ref{fig:FigCutPassivatedSetup}A).  A closed loop of desired size is then identified so as to intersect single bonds only.  The carbon atoms within the closed loop are then removed and the remaining carbons with missing neighbors become the binding sites for the passivating hydrogens  (Figure ~\ref{fig:FigCutPassivatedSetup}B). 
It is necessary to note that in reality, double bonds are delocalized on a graphene sheet. There are therefore multiple ways of drawing the double bonds with each decoration leading to its own set of possible shapes and sizes.

After generating the passivated pore, the positions of hydrogen atoms are relaxed in a density functional theory (\textsc{Dft}) simulation using \textsc{Gaussian 16}\cite{g16} with the \textsc{B3lyp} functional\cite{TiradoRivesJChemTheoryComput2008} and an \textsc{Sto-3g} basis set.\cite{HehreJChemPhys1969} For the present membrane structure, the differences between initial and relaxed positions of the hydrogen atoms are negligible. The relaxed positions are then used to construct a three-layer graphite membrane perpendicular to the $z$ axis. Graphene pistons are constructed using \textsc{Vmd}\cite{HumphreyJMolGraphics1996} and are placed on either side of the membrane. Finally, \textsc{Packmol}\cite{AlloucheJComputChem2012} is used for adding sodium and chloride ions and water molecules to the feed and water molecules to the filtrate to generate 100 independent starting configurations. A representative configuration generated using this procedure is shown in Fig.~\ref{fig:FigCutPassivatedSetup}C. In order to examine the impact of system size on the kinetics of water and ion transport, this procedure is repeated to construct systems with identical ionic concentrations and pore geometries but with different box dimensions along at the $x$ and $y$ directions.  In particular, we examine systems with membrane cross-sectional areas of $12.53$~nm$^{2}$, $26.33$~nm$^{2}$, $32.60$~nm$^{2}$, $50.15$~nm$^{2}$ and $100.28$~nm$^{2}$ in this work, with the correstponding box dimensions and the number of water molecules and ion pairs given in Table~\ref{tab:system}.

\subsection{NEMD Simulations and System Equilibration}
\label{section:methods:NEMD}
 
\noindent 
Each independent starting configuration is equilibrated using \textsc{Lammps},\cite{PlimptonJComputPhys1997} an open source \textsc{Md} simulation engine. All atoms are represented as charged Lennard-Jones (\textsc{Lj}) particles, with water molecules and sodium and chloride ions modeled using the \textsc{Tip3p}\cite{PriceJChemPhys2004} and Joung-Cheatham (\textsc{Jc})\cite{JoungJPhysChemB2008} force-fields, respectively. The \textsc{Lj} parameters for carbons and passivating hydrogens were adopted from M\"{u}ller-Plath\cite{MullerMacromolecules1996} and Beu\cite{BeuJChemPhys2010}. All interaction parameters, partial charges and cutoff distances are given in our earlier publication.\cite{MalmirMatter2020} All \textsc{Md} trajectories are carried out using the velocity-Verlet algorithm\cite{SwopeJChemPhys1982} while temperature is maintained at 300~K using the Nos\'{e}-Hoover thermostat\cite{NoseMolPhys1984, HooverPhysRevA1985, EvansJChemPhys1985} with a damping time constant of 0.1~ps.  The \textsc{Shake} algorithm\cite{AndersenJComputPhys1983} is used to maintain the rigidity of the water molecules.   Long-range electrostatic interactions are estimated using the  particle-particle particle-mesh (\textsc{Pppm}) method\cite{Hockney_1988} with a short-range cutoff of 1~nm. Periodic  boundary conditions are only applied along the $x$ and $y$ directions in order to avoid artifacts due to unphysical long-range electrostatic interactions between the system and its periodic images along the $z$ axis.\cite{BostickBiophysJ2003} We therefore treat long-range interactions using a modified version of the \textsc{Pppm} method known as slab \textsc{Pppm}. 

\begin{table}
	\vspace{-5pt}
	\caption{Summary of the parameters utilized for system setup, including $L_x$ and $L_y$, the dimensions of the simulation box along the $x$ and $y$ directions, $N_{w,\text{feed}}$ and $N_{w,\text{filtrate}}$, the number of water molecules added to the feed and filtrate compartments and $N_{i,\text{feed}}$, the number of ion pairs added to the feed.
	\label{tab:system}}
\begin{tabular}{ccc|c|c|c}
\hline
\hline
$L_x~[\text{nm}]$ & $L_y~[\text{nm}]$ & Area~[nm$^2$] & $N_{w,\text{feed}}$ & $N_{w,\text{filtrate}}$ & $N_{i,\text{feed}}$  \\
\hline 
~3.684 & ~3.403 & ~12.537 & $~1,616$ & $1,095$ & $~45$ \\
~5.158 & ~5.105 & ~26.329 & $~3,393$ & $2,300$ & $~95$ \\
~6.386 & ~5.105 & ~32.598 & $~4,220$ & $2,857$ & $118$\\
~7.614 & ~6.806 & ~51.822 & $~6,473$ & $4,380$ & $180$\\
10.070 & 10.210 & 102.808 & $~13,275$ & $8,982$ & $371$\\
\hline
\end{tabular}
\vspace{-10pt}
\end{table}

Similar to our earlier work, we use an \textsc{Nemd} scheme in which the same force is exerted on all atoms within each piston in order to maintain a net pressure gradient of 196.6 bar across the membrane. For this purpose, we use the \texttt{fix aveforce} keyword in \textsc{Lammps}. During initial equilibration, the system is simulated for 10~ps with a time step of 0.5~fs while the pistons are kept fixed. We then allow the pistons to move according to the scheme described above, and equilibrate the system for an additional 2~ns with a time step of 1~fs. The endpoints of these trajectories are utilized for the j\textsc{Ffs} calculations described below wherein all trial trajectories are integrated using a time step of 1~fs.

\subsection{Path Sampling Calculations}
\label{section:methods:ffs}

\noindent
We compute  ionic passage times using the forward-flux sampling\cite{Allen2006} (\textsc{Ffs}) method,  an advanced path sampling technique that allows for efficient and accurate characterizations of rare events.\cite{HussainJChemPhys2020} One of the main advantages of \textsc{Ffs} is its compatibility with microscopic irreversibility. As such, in addition to equilibrium processes such as crystal nucleation\cite{ValeriniJChemPhys005, LiNatMater2009, ThaparPRL2014, HajiAkbariPCCP2014, HajiAkbariPNAS2015, GianettiPCCP2016,  HajiAkbariPNAS2017, JiangJChemPhys2018}, hydrophobic evaporation\cite{SumitPNAS2012, AltabetPNAS2017} and protein folding,\cite{BorreroJCP2006} \textsc{Ffs} has been recently used for probing non-equilibrium processes such as phase separation\cite{RichardSoftMatter2016} and morphological changes\cite{LiYaoPRL2019} in active matter systems, protein rupture\cite{GoncharovPRL2005} and polymer translocation,\cite{HernandezOrtizJChemPhys2009} and it is a remarkably powerful tool for probing the kinetics of pressure-driven ion transport. Similar to other path sampling techniques, \textsc{Ffs} requires constructing an order parameter, $\lambda{(\cdot)}$, a mechanical observable that 
is a measure of progress towards the completion of the rare event under consideration. In the case of ion passage through a rigid membrane, we choose $\lambda(\cdot)$ to be the curved directed distance of the leading ion from the pore opening with larger $\lambda$ values corresponding to a leading ion which has traversed farther into the pore. The precise mathematical definition of $\lambda(\cdot)$ for our specific pore geometry is given in our earlier work.\cite{MalmirMatter2020}  In general, \textsc{Ffs} characterizes the kinetics of transition between two (meta)stable basins that are demarcated by $\lambda(\cdot)$,~i.e.,~ ${A:=\{x\in\mathcal{Q}:}\lambda{(x)} < \lambda_A{\}}$ and ${B:=\{x\in\mathcal{Q}:}\lambda{(x)}\ge\lambda_B{\}}$ and that are separated by large kinetic barriers. (Here, $\mathcal{Q}$ is the \emph{configuration space} that contains all configurational degrees of freedom of the corresponding system.) In the case of membrane transport, these two basins are denoted by $F_{0,0}$ and $F_{0,1}\cup F_{1,0}$ where $F_{p,q}$ signifies the set of configurations in which $p$ sodium and $q$ chloride ions  are present in the filtrate.  While this OP does not distinguish between the sodium and chloride ions, the leading ion traversing the pore is always a chloride, due to the partial charge of the passivating hydrogens within the pore. Our approach therefore effectively probes the $F_{0,0}\rightarrow F_{0,1}$ transition consistent with our earlier work.\cite{MalmirMatter2020} 
		
Upon identifying a proper order parameter, an \textsc{Ffs} calculation is initiated by sampling the starting basin (in this case, $F_{0,0}$) using unbiased \textsc{Nemd} trajectories launched from the independent configurations generated within $A$.  After equilibration, a long \textsc{Nemd} trajectory is launched from each configuration and is monitored for first crossings of $\lambda_0$ every time it leaves $A$. The configurations corresponding to such crossings are saved for future iterations. Here, $\lambda_0(>\lambda_A)$, is a milestone sufficiently close to $\lambda_A$ so that it can be reached frequently enough by trajectories initiated within $A$.  These trajectories are terminated upon sufficient sampling of $A$, and the initial $\Phi_0$, which corresponds to the number of first crossings of $\lambda_0$ per unit time, is estimated as,
\begin{eqnarray}
{\Phi_0:=\frac{N_0}{T}}
\end{eqnarray}
Here, $N_0$ is the total number of successful crossings of $\lambda_0$ while $T$ is the total length of \textsc{Nemd} trajectories. 
	
\begin{figure*}
	\centering
	\includegraphics[width=0.7965\textwidth]{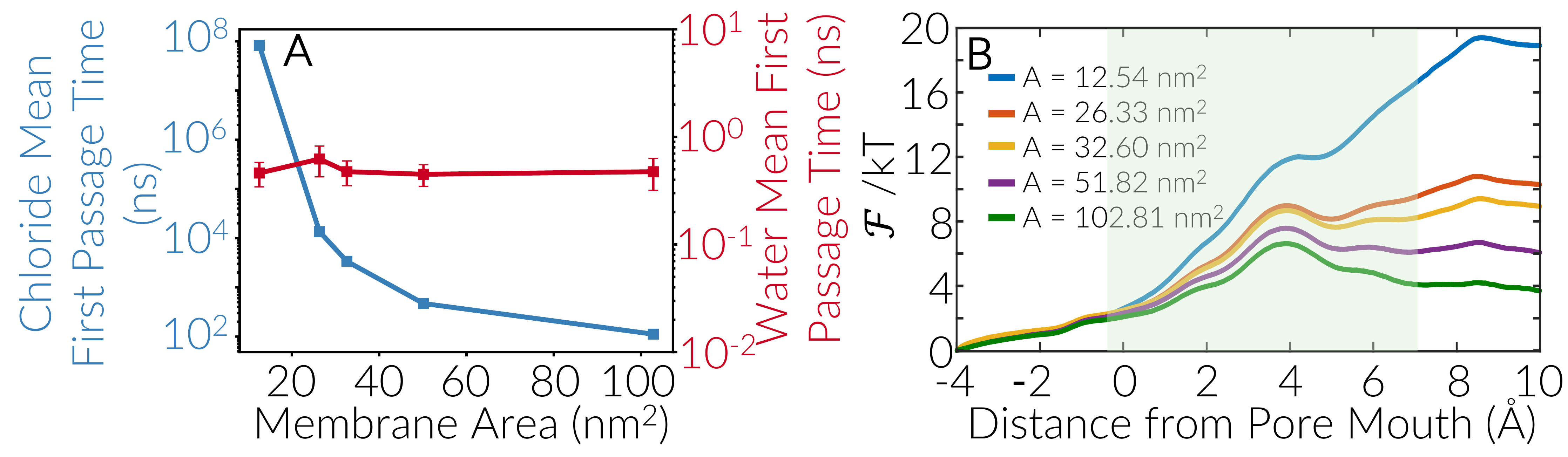}
	\vspace{-10pt}
	\caption{\label{fig:FigPassageTimeFreeEnergy}(A) Mean first passages time for chloride ions and water molecules and (B) generalized free energy profiles for different system sizes.  The shaded region in (B) corresponds to the pore interior. System size impacts both the magnitude of the free energy barrier and the locus of the transition state.}
	\vspace{-10pt}
\end{figure*}

		After finishing the sampling of the starting basin, another target milestone, $\lambda_1$, is specified. As outlined in an earlier publication\cite{HajiAkbariJChemPhys2018}, $\lambda_1$ should be placed beyond $\lambda_{0,\max}$, or the largest value of the order parameter for configurations obtained after first crossing of $\lambda_0$. Trial trajectories are then initiated from the stored configurations at or beyond $\lambda_0$ in which the velocities of water molecules and sodium and chloride ions are randomized according to the Boltzmann distribution. These trajectories are continued until they reach $\lambda_1$ or return to the starting basin. The configurations corresponding to successful crossings of $\lambda_1$ are saved. After sufficient sampling, the transition probability from $\lambda_0$ to $\lambda_1$,  $P(\lambda_1|\lambda_0)$, is computed by dividing the number of trajectories resulting in a successful crossing of $\lambda_1$ by the total number of trial trajectories. This process is repeated iteratively to compute transition probabilities between successive milestone. Once all the transition probabilities are known, the total rate is computed as
\begin{equation*}
\Phi=\Phi_0\prod_{k=0}^{N-1} P(\lambda_{k+1}|\lambda_k)\,.
\end{equation*}
Note that $\tau$, the mean first passage time for a trajectory initiated in $A$ to reach $B$ is then estimated as $\tau=1/\Phi$. 

\allowdisplaybreaks

\subsection{Generalized Free Energy Profiles}
\label{section:methods:ffs-mfpt}

\noindent
The notion of a Landau free energy $\mathcal{F}(\textbf{q})$ over a collective variable space $\textbf{q}$ can only be defined unambiguously for systems with stationary probability distributions.\cite{HussainJChemPhys2020} Strictly speaking, this condition is not satisfied for non-equilibrium processes such as pressure-driven solute transport. However, one might still be able to define a generalized notion of free energy as long as the system is under pseudo-steady state conditions. This is generally true for most non-equilibrium processes with long induction times. Here, we assume the existence of a notion of a generalized free energy, and compute it using the forward-flux sampling/mean first passage time (\textsc{Ffs}-\textsc{Mfpt}) method.\cite{ThaparJCP2015} Considering the diffusive nature of the leading ion's motion, we can use the \textsc{Ffs}-\textsc{Mfpt} method only with minor adjustments to account for apparent jumpiness in OP due to discretization. In particular, we estimate $\tau(\lambda;A)$, the mean  first passage time for reaching $\lambda\in(\lambda_k,\lambda_{k+1}]$ from A, as
\begin{eqnarray}
\tau(\lambda;A) = \frac{1}
{M_k^{\lambda^+}F_k^{\lambda^+}+{S_\lambda}\left(1-F_k^{\lambda^+}\right)}\Bigg\{
 M_k^{\lambda^+}\tau(\lambda_k;A)\notag\\
 + \left(1-F_{k}^{\lambda^+}\right)
 \left[
L_\lambda^{(s)}{S_\lambda}
+L_{\lambda}^{(f)}({M_k^{\lambda^+}-S_\lambda})\right]\Bigg\}
\label{eq:MFPT}
\end{eqnarray}
Here, $F_k^{\lambda^+}$ is the fraction of crossing events at $\lambda_k$ that land at a configuration with an order parameter value $\ge\lambda$, $M_k^{\lambda^+}$ is the total number of trial trajectories during the $(k+1)$th iteration launched from configurations with an order parameter value $<\lambda$, $S_\lambda$ is the number of those trajectories that reach $\lambda$, and $L^s_\lambda$ and $L^f_\lambda$ are the average durations of the $S_\lambda$ successful and $M_k^{\lambda^+}-S_\lambda$ failing trajectories, respectively. The derivation of Eq.~(\ref{eq:MFPT}) is given in Appendix~\ref{secction:jumpy-correction}.

\section{Results}

\noindent
Figure ~\ref{fig:FigPassageTimeFreeEnergy}A depicts $\tau_w$ and $\tau_c$, the mean first passage times for water molecules and chloride ions computed from conventional \textsc{Nemd} and j\textsc{Ffs} calculations, respectively. While $\tau_w$ is virtually insensitive to system size, $\tau_c$ exhibits a strong dependence on system size and varies by almost six orders of magnitude for the range of system sizes considered in this work. We also use the \textsc{Ffs}-\textsc{Mfpt} method to estimate $\mathcal{F}(\lambda)$, the generalized free energy profiles as a function of the order parameter, with the computed $\mathcal{F}(\lambda)$ profiles depicted in Fig.~\ref{fig:FigPassageTimeFreeEnergy}B. Remarkably, the sensitivity of $\mathcal{F}(\lambda)$ to system size goes well beyond the magnitude of the free energy barrier, and extends to the number and loci of the free energy maxima. More precisely, $\mathcal{F}(\lambda)$ exhibits two major peaks for smaller systems with the larger peak-- and the main barrier to transport-- located right after the pore exit. As the system size increases, however, the second peak weakens and eventually disappears. Therefore, the locus of the transition state moves discontinuously into the pore interior for sufficiently large system sizes. This dramatic change in the qualitative features of the translocation mechanism implies that examining a system that is too small will not only lead to an underestimation of the ionic flux, but will also result in inaccurate identification of the ion transport mechanism.

\begin{figure}
	\centering
	\includegraphics[width=0.3308\textwidth]{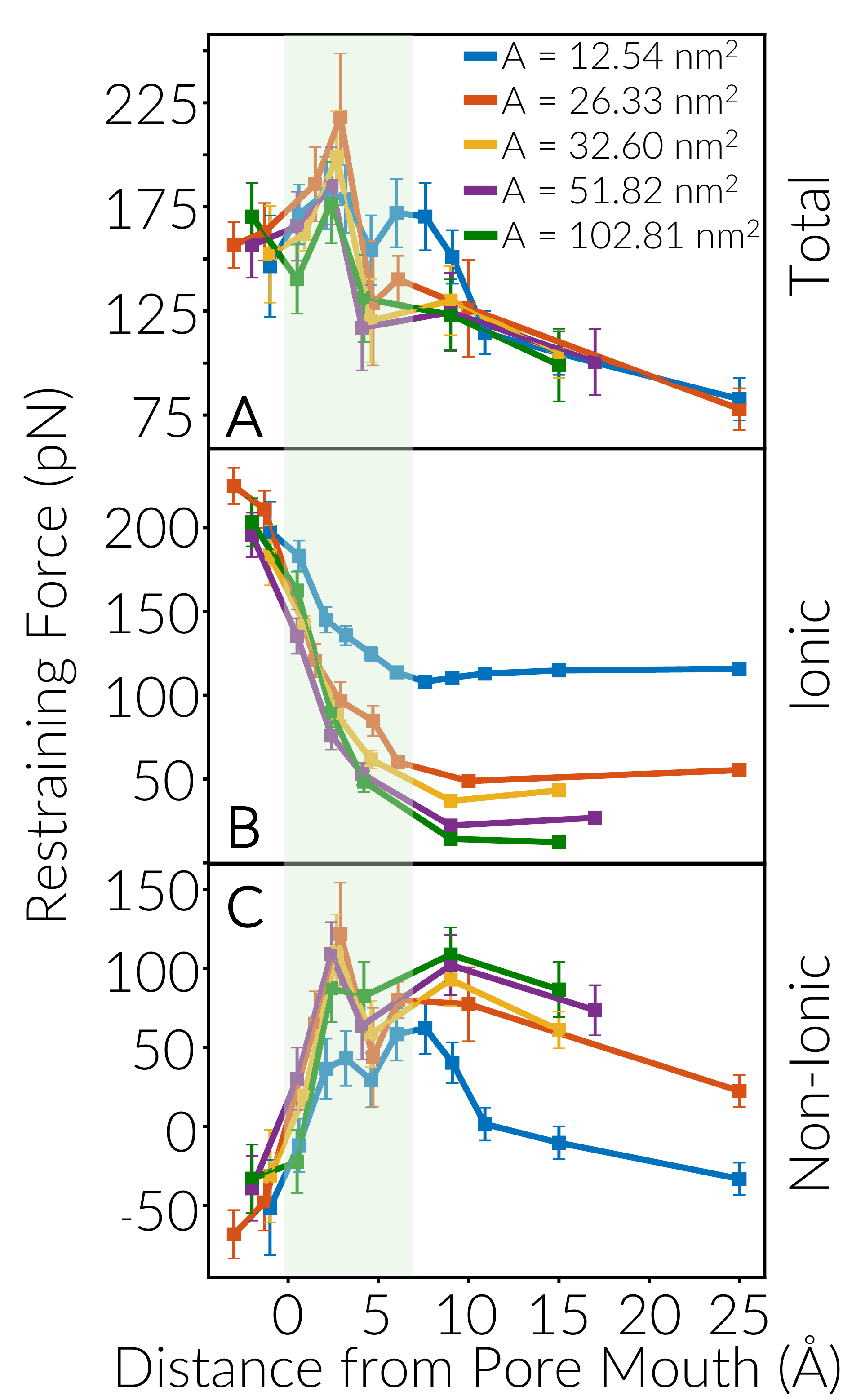}
	\vspace{-10pt}
	\caption{\label{fig:FigForces} (A) $\langle -f_z\rangle$, the magnitude of the restraining force exerted on the leading ion at different \textsc{Ffs} milestones, as well as its breakdown to its (B) ionic and (C) non-ionic contributions.}
	\vspace{-20pt}
\end{figure}

In order to understand the origin of this strong sensitivity, we first compute $\langle -f_z\rangle$, the $z$ component of the average restraining force exerted on the leading ion. As discussed in our earlier publication,\cite{MalmirMatter2020} $\langle -f_z\rangle$ is non-vanishing even after the ion has left the pore, and its magnitude is determined by an interplay among ion-pore and ion-water interactions, the hydrodynamic resistance to ion mobility, and the charge anisotropy induced upon the passage of the leading ion.  
As can be seen in Fig.~\ref{fig:FigForces}A, there is no noticeable trend for the dependence of $\langle -f_z\rangle$ on system size. A clear trend emerges, however, upon decomposing $\langle -f_z\rangle$ into its ionic (Fig.~\ref{fig:FigForces}B) and non-ionic (Fig.~\ref{fig:FigForces}C) contributions. More precisely, $\langle -f_{z,\text{ionic}}\rangle$, the force exerted on the leading ion by other ions in the system, is stronger in smaller systems and diminishes in magnitude upon increasing the system size.  As discussed in detail in our earlier work,\cite{MalmirMatter2020} $\langle -f_{z,\text{ionic}}\rangle$ primarily arises due to the charge anisotropy induced by the leading ion within the feed. More precisely, when a chloride enters the pore, it induces a net charge surplus $+\text{\calligra\Large e}\,$ within the feed, which then exerts a restraining force on it. The observed differences in the ionic restraining force is therefore likely caused by how the distribution of the surplus charge is impacted by system size.

\begin{figure*}
	\centering
	\includegraphics[width=.9653\textwidth]{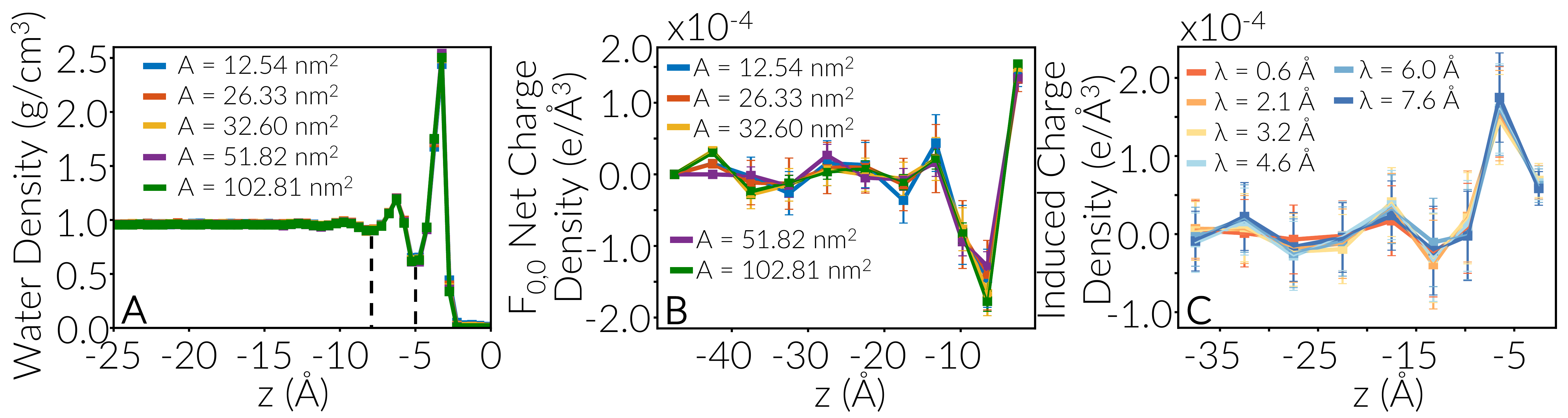}
	\caption{\label{fig:FigDensity} (A) Water and (B)  net charge density within the feed compartment as a function of $z$, the distance from the membrane surface, in the $F_{0,0}$ basin,~i.e.,~before any charge anisotropy is established. Water density is computed using cuboidal bins of thickness 0.25~{\AA} while the binning scheme for the net charge density is discussed in the text. (C) The induced charge density in the system with the 12.53~nm$^{2}$ cross-sectional area, demonstrating  that the induced charge is distributed in the first two liquid layers at the membrane surface.}
\end{figure*}

In order to determine how this excess charge is distributed within the feed, it is first necessary to characterize the intrinsic charge distribution within the feed \emph{prior to} the leading chloride's entrance into the pore. This is achieved by partitioning the feed into cuboidal bins and computing the charge density within each bin. Considering the structuring of water molecules at the membrane surface, we demarcate the first few bins using the valleys of  $\rho_w(z)$, the water density as a function of $z$, the distance from the membrane surface (Fig.~\ref{fig:FigDensity}A). Within the plateau region of $\rho_w(z)$, however, a uniform bin width of 0.5~nm is utilized. Fig.~\ref{fig:FigDensity}B depicts the computed charge density profiles within $F_{0,0}$,~i.e.,~when all the ions are in the feed. Notably, a non-uniform charge distribution-- a
double layer in particular-- is established at the membrane surface due to its overall affinity towards sodium ions. More precisely, the radially outward dipoles within the pore interior result in an affinity towards chlorides right at the pore entrance,  but  sodiums at the rest of the membrane surface.
Considering the meager surface area of the pore opening in comparison to the rest of the membrane, the first layer of the liquid is populated by sodium ions overall, resulting in a positive charge density, followed by two layers dominated by chloride ions and  a negative charge density. Note that these intrinsic profiles are independent of system size suggesting that the strong sensitivity of rate and $\langle -f_{z,\text{ionic}}\rangle$ to system size is likely caused by the induced charges.

We then compute the \emph{induced} charge density profiles by estimating the net charge profiles at different \textsc{Ffs} milestones and subtracting from them the intrinsic profiles of Fig.~\ref{fig:FigDensity}B. Fig.~\ref{fig:FigDensity}C depicts the induced profiles for the smallest system. We observe that the induced charge resides entirely within the first two layers of water at the membrane surface. The accumulation of the excess charge at the membrane surface is qualitatively consistent with the feed compartment being an ideal conductor. From a physical perspective, this is a reasonable approximation since the sodium and chloride ions can act as mobile charge carriers that can redistribute until the net electric field in the bulk vanishes.

\begin{figure}
	\centering
	\includegraphics[width=0.3423\textwidth]{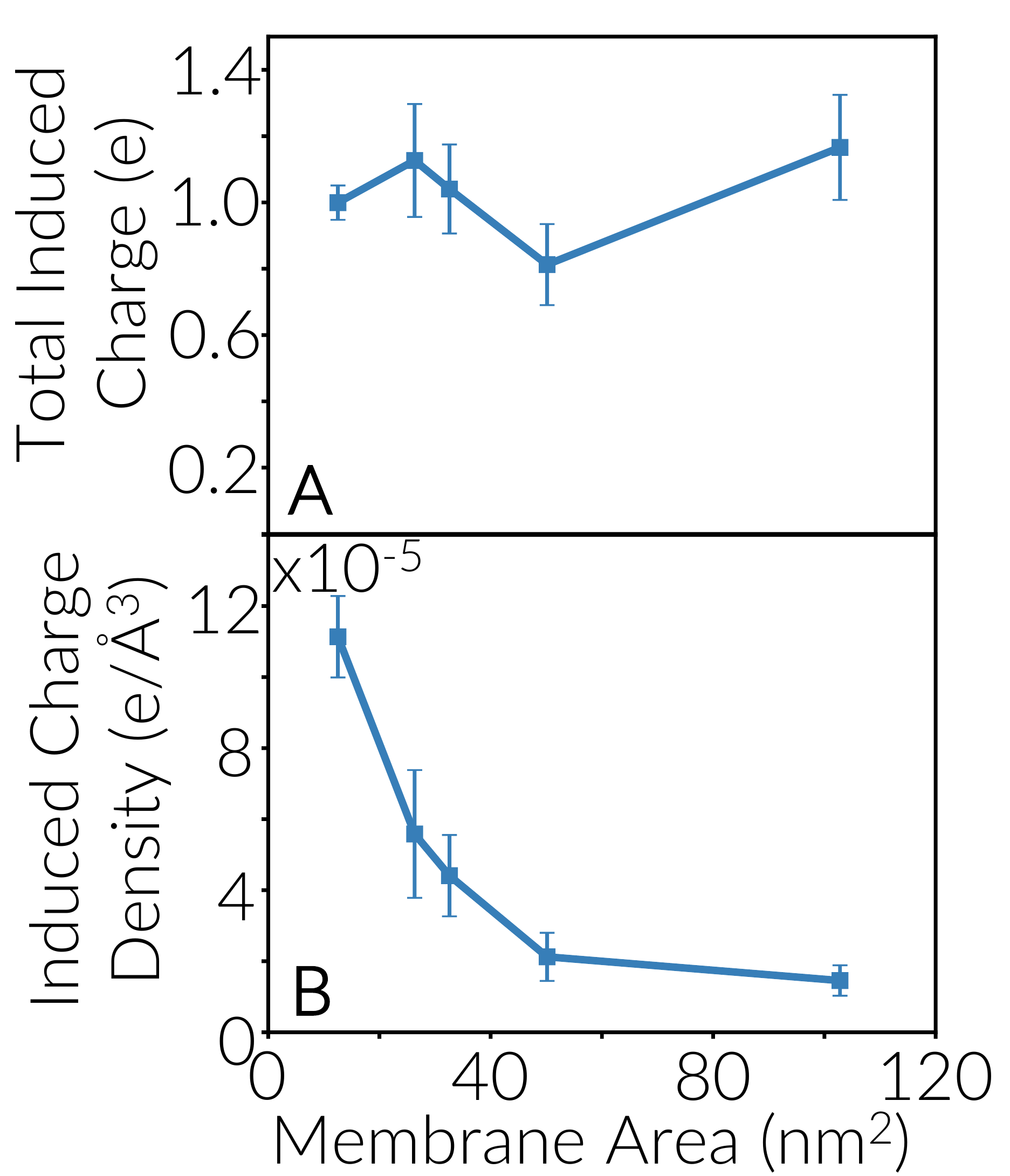}
	\vspace{-10pt}
	\caption{\label{fig:FigInducedCharge}(A) The total induced charge and (B) the induced charge density contained within the first two liquid layers at the membrane surface vs.~system size. Irrespective of system size, an induced charge of $\approx+\text{\Large\calligra e}\,$ is contained within the first two layers. } 
	\vspace{-15pt}
\end{figure}

The accumulation of excess charge at the membrane surface is observed for all system sizes (SI Fig.~S1). As can be seen in Fig.~\ref{fig:FigInducedCharge}A, the total induced charge is fully contained within the first two layers of water at the membrane surface irrespective of system size, which is the expected behavior if the feed acts as a  conducting slab. Considering the larger surface area of the membrane in larger systems,  this net excess charge is distributed over a larger area, which will lead to lower charge densities (Fig.~\ref{fig:FigInducedCharge}B) and likely smaller ionic restraining forces. 

\allowdisplaybreaks

\begin{table}
	\vspace{-5pt}
	\caption{The location of the surface of the feed conductor, $z_c$, in the ideal conductor model. All uncertainties correspond to 95\% confidence intervals.
	\label{tab:charge-center}}
\begin{tabular}{c|c}
\hline
\hline
 ~~~~Area~[nm$^2$]~~~~ & ~~~~$z_c$~[nm]~~~~  \\
\hline 
 ~~~~~12.537~~~~ & ~~~~$-0.464\pm0.014$~~~~ \\
 ~~~~~26.329~~~~ & ~~~~$-0.430\pm0.054$~~~~ \\
~~~~~32.598~~~~ & ~~~~$-0.443\pm0.044$~~~~\\
~~~~~51.822~~~~ & ~~~~$-0.421\pm0.046$~~~~\\
~~~~102.808~~~~ & ~~~~$-0.412\pm0.040$~~~~\\
\hline
\end{tabular}
\vspace{-15pt}
\end{table}

In order to determine whether our computed rates and generalized free energy profiles are \emph{quantitatively} consistent with the feed compartment being an ideal conductor, we use the tools of electrostatics to calculate the lateral density of the  accumulated excess charge, as well as the ensuing  distortions in free energy. The charge density accumulated at the surface of an infinitely large conducting slab upon the removal of a monovalent anion from it can be estimated using the method of images and is given by:
\begin{eqnarray}
\rho_\infty(x,y) &=& \frac{\text{\calligra\Large e}\,h}{2\pi\left[x^2+y^2+h^2\right]^{3/2}}.
\label{eq:rho-inf}
\end{eqnarray}
Here, $h$ is the distance of the anion from the surface of the slab. Note that $\rho_\infty(x,y)$ can be used for estimating the contribution of induced charge anisotropy to $\mathcal{F}(\lambda)$ in the thermodynamic limit. The observed dependence of $\tau_c$ and $\Delta\mathcal{F}^*$ on system size, however, cannot be characterized using Eq.~(\ref{eq:rho-inf}). This is because removing an anion from the feed compartment in a simulation box that is periodic along the $x$ and $y$ dimensions does result in the removal of all its periodic images, and each such periodic image will induce its own charge density within the slab. The ensuing total charge density accumulated at the membrane surface can be readily estimated using the superposition principle (See Appendix~\ref{section:correction}) and $\widetilde{\rho}_f(x,y)=L_xL_y\rho_f(x,y)/\text{\calligra\Large e}\,$, the dimensionless charge density for the periodic system,  will be given by,
\begin{eqnarray}
\widetilde{\rho}_f(x,y) &=& 1 + 2\sum_{\alpha\in\{x,y\}}\sum_{k_\alpha=1}^{+\infty} e^{-q_\alpha h} \cos q_\alpha\alpha+\notag\\
&& 2\sum_{k_x,k_y=1}^{+\infty}e^{-|\textbf{q}|h}\sum_{p=\pm1}
\cos (q_xx+pq_yy)
\label{eq:main-superposition-reciprocal}
\end{eqnarray}
wherein $\textbf{q}=(2\pi k_x/L_x, 2\pi k_y/L_y)\in\mathbb{R}^2$ is the wavevector associated with $(k_x,k_y)\in\mathbb{N}^2$. 
While we cannot directly test the validity of Eq.~(\ref{eq:main-superposition-reciprocal}) due to lack of sufficient statistics, we can use Eq.~(\ref{eq:main-superposition-reciprocal}) to determine the excess increase in $\mathcal{F}(\lambda)$ due to $\Delta\rho(x,y)=\rho_f(x,y)-\rho_\infty(x,y)$. This is achieved by estimating the \emph{excess} electric field induced by $\Delta\rho(x,y)$ and integrating it to obtain the following \emph{correction} to the free energy:
\begin{eqnarray}
\frac{\Delta\mathcal{F}_{\text{corr}} (z)}{kT} 
&=& \frac{\text{\calligra\Large e}\,^2}{2\epsilon_0kTL_xL_y}\Bigg\{
z-z_0\notag\\
&&-\frac{L_xL_y}{2\pi}\left[
\frac{1}{z_0+z}-\frac{1}{2z_0}
\right]\notag\\
&& -~2\sum_{\alpha\in\{x,y\}}\sum_{k_\alpha=1}^{+\infty}  \frac{e^{-2q_\alpha z}-e^{-q_\alpha(z+z_0)}}{q_\alpha}
\notag\\
&& -~4\sum_{k_x,k_y=1}^{+\infty} \frac{e^{-2|\textbf{q}|z}-e^{-|\textbf{q}|(z+z_0)}}{|\textbf{q}|}
\Bigg\}
\label{eq:fe-corr}
\end{eqnarray}
Here, $\epsilon_0$ and $k$ are the vacuum permittivity and the Boltzmann constant, respectively, and $z_0$ is the distance from the conductor surface at which the anion is fully detached from the feed and the induced charge density is established. It is  not trivial to rigorously define what constitutes `full detachment' considering the connectivity of the leading ion to the feed through a train of water molecules. One can, however, argue that the leading chloride becomes fully detached when its first hydration shell has completely left the first liquid layer at the membrane surface. This will correspond to $\lambda\approx0$,~i.e.,~when the ion starts entering the pore. This assertion is corroborated by our observation that the $\mathcal{F}(\lambda)$ profiles start diverging at $\lambda\approx0$ as can be seen in Fig.~\ref{fig:FigPassageTimeFreeEnergy}B.

Prior to applying this correction to the free energy profiles depicted in Fig.~\ref{fig:FigPassageTimeFreeEnergy}B, two additional points need to be addressed. First,  the free energy correction given in Eq.~(\ref{eq:fe-corr}) is a function of $z$, which differs from the curvilinear order parameter utilized in \textsc{Ffs} and  only aligns with it within the pore interior. Since the size-dependent uphill portion of $\mathcal{F}(\lambda)$ is mostly contained within the pore interior, any errors due to such mismatch will be minimal (and likely inconsequential).  The second issue is that both $z$ and $z_0$ correspond to the distance from the surface of the conductor which does not necessarily coincide with the membrane surface.
While such a distinction might be unimportant at the macroscopic scale, it can become significant at a molecular level. Moreover, the region in which the induced excess charge is distributed will be diffuse, which will make it nontrivial to define what constitutes the conductor surface. Considering our observation that the excess charge is almost fully contained within the first two layers of water at the immediate vicinity of the membrane, we define $z_c$, the locus of the conductor surface, as the geometric center of the induced charges within the first two liquid layers. More precisely, we compute $z_c$ as $z_c=(z_1q_{i,1}+z_2q_{i,2})/(q_{i,1}+q_{i,2})$ wherein $z_j$ and $q_{i,j}$ are the average $z$ coordinate of and the total induced charge contained within the $j$th liquid layer, respectively. Using this definitions, we obtain $z_c$ values that are for the most part independent of the system size and vary between $-0.46$ and $-0.41$~nm as can be seein in Table~\ref{tab:charge-center}.

\begin{figure*}
	\centering
	\includegraphics[width=0.6876\textwidth]{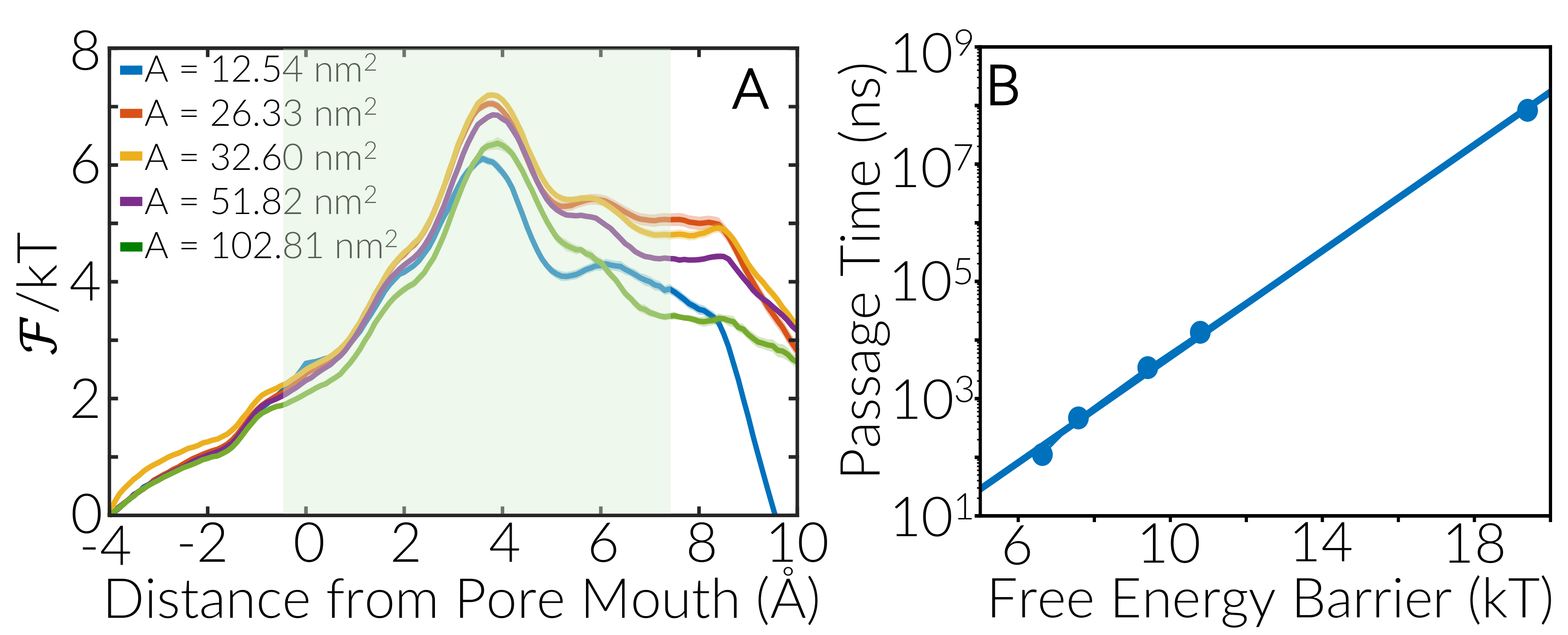}
	\caption{\label{fig:FigEnergyBarrierPassageTime}(A) The corrected generalized free energy profiles for different system sizes after applying the ideal conductor model. (B) A linear relationship between $\Delta\mathcal{F}^*_f$ and $\ln\tau_{c,f}$.} 
\end{figure*}

After estimating the model parameters, we subtract the finite size correction given by Eq.~(\ref{eq:fe-corr}) from the $\mathcal{F}(\lambda)$ profiles depicted in Fig.~\ref{fig:FigPassageTimeFreeEnergy}B to compute  $\mathcal{F}_{\infty}(\lambda)\equiv \mathcal{F}_f(\lambda)-\mathcal{F}_{\text{corr}}(\lambda)$, the generalized free energy profile in the thermodynamic limit. The corrected free energy profiles  are depicted in Fig.~\ref{fig:FigEnergyBarrierPassageTime}A. While each profile has been  estimated independently,~i.e.,~only using the information obtained from the corresponding finite size simulation, they are all in excellent agreement with one another both in terms of the loci of the transition state and the magnitudes of the free energy barrier. Overall, they predict a transition state that is located at $\lambda^*\approx0.38\pm0.02$~nm and has a free energy barrier of $\Delta\mathcal{F}_\infty(\lambda^*)/kT=6.82\pm0.12$. The barriers obtained from the individual profiles are also fairly consistent and vary between $6.1kT$ and $7.2kT$. This is in stark contrast to the original (uncorrected) free energy profiles that exhibit substantial differences not only in terms of the locus of the transition state but also the magnitude of the free energy barrier, the latter varying by $13kT$.

The next step is to estimate $\tau_c$ in the thermodynamic limit. While it is in principle possible to relate the free energy profile to the passage time using the reaction rate theory,\cite{PetersRxnRate2017} doing so requires estimating $D(\lambda)$, or diffusivity along the $\mathcal{F}_f(\lambda)$ and $\mathcal{F}_\infty(\lambda)$ free energy landscapes. For sufficiently large free energy barriers, however, it can be demonstrated that the passage time will be given by $\tau=\mathcal{A}\exp\left[+\Delta\mathcal{F}^*/kT\right]$ wherein $\mathcal{A}$ is a prefactor that depends on, among other things, the local curvatures of the starting basin and the transition state.\cite{KramersPhysicaA1940, LangerAnnPhys1969} Intuitively, this expression is based on the assumption that a quasi-equilibrium is established between the starting basin and the transition state. We argue that $\mathcal{A}$ has to be  insensitive to system size. This assertion is motivated by the observed mechanism of finite size scaling that kicks in when the ion fully enters the pore, and is bolstered by the insensitivity of properties such as $\tau_w$ and the pre-entrance portion of $\mathcal{F}(\lambda)$ to system size. Moreover, the validity of this assertion is unambiguously corroborated by the  perfect linear scaling  between $\ln\tau_c$ and $\Delta\mathcal{F}^*$ ($R^2=0.99$) as depicted in  Fig.~\ref{fig:FigEnergyBarrierPassageTime}B. It is therefore reasonable to argue that $\tau_{c,\infty}$ can be estimated as
\begin{eqnarray}
\tau_{c,\infty} &=& \tau_{c,f} \exp\left[
\frac{\Delta\mathcal{F}^*_\infty-\Delta\mathcal{F}^*_f}{kT}
\right].\label{eq:Arrhenius}
\end{eqnarray}
Eq.~(\ref{eq:Arrhenius}) can be used to estimate $\tau_c$ in the thermodynamic limit from the information obtained from each simulation, with the corresponding estimates given in Table~\ref{table:PassageTimes}. While these $\tau_{c,\infty}$'s have been estimated independently, they only differ by a factor of four despite the astronomical differences among the $\tau_{c}$'s computed for finite systems. A  holistic estimate can, however, be obtained by using the linear correlation depicted in Fig.~\ref{fig:FigEnergyBarrierPassageTime}B. More precisely, a $\Delta\mathcal{F}^*_\infty/kT=6.82\pm0.12$ yields a passage time of $\log_{10}\tau_{c,\infty}~[ns]=2.35\pm0.45$ (or $\tau_{c,\infty}=224_{-144}^{+407}$~ns), which is consistent with the individual estimates given in Table~\ref{table:PassageTimes}.

\begin{table*}
\centering
\caption{\label{table:PassageTimes}Estimates of $\tau_c$ and $\Delta\mathcal{F}^*/kT$ from j\textsc{Ffs} and extrapolations to the thermodynamic limit using the ideal conductor model. All uncertainties correspond to 95\% confidence intervals. }
\begin{tabular}{cccccccc}
\hline
\hline
~~$L_x~[\text{nm}]$~~ & ~~$L_y~[\text{nm}]$~~ & ~~Area~[nm$^2$]~~ & ~~$\tau_{c,f}~[\text{ns}]$~~ & ~~$\Delta\mathcal{F}^*_f/kT$~~ & ~~$\Delta\mathcal{F}^*_\infty/kT$~~ & ~~$
\tau_{c,\infty}~[\text{ns}]$ &~~  \\
\hline 
~~~3.684~~ & ~~~3.403~~ & ~~~12.537~~ & ~~$(8.29\pm0.53)\times10^7$~~ & ~~$19.40\pm0.10$~~ & ~~$6.11\pm0.12$~~ & ~~$140\pm48$~~\\
~~~5.158~~ & ~~~5.105~~ & ~~~26.329~~ & ~~$(1.36\pm0.17)\times10^4$~~ & ~~$10.79\pm0.06$~~ & ~~$7.05\pm0.14$~~ & ~~$322\pm117$~\\
~~~6.386~~ & ~~~5.105~~ & ~~~32.598~~ & ~~$(3.38\pm0.17)\times10^3$~~ & ~~~$9.41\pm0.04$~~ & ~~$7.20\pm0.14$~~ & ~~$372\pm79$~~\\
~~~7.614~~ & ~~~6.806~~ & ~~~51.822~~ & ~~$(4.69\pm0.58)\times10^2$~~ & ~~~$7.58\pm0.08$~~ & ~~$6.86\pm0.10$~~ & ~~$228\pm68$~~\\
~~10.070~~ & ~~10.210~~ & ~~102.808~~ & ~~$(1.12\pm0.16)\times10^2$~~ & ~~~$6.63\pm0.22$~~ & ~~$6.37\pm0.22$~~ & ~~~$86\pm64$~~\\
\hline
\end{tabular}
\end{table*}

We finally examine the sensitivity of the conclusions from our earlier work\cite{MalmirMatter2020}  to the finite size effects discussed here. While the box dimensions  in Ref.~\citenum{MalmirMatter2020} match those of the second smallest system in this work, the pore chemistries are slightly different as discussed in Section~\ref{section:methods:setup}. Such differences not only affect the kinetics and mechanism of ion transport, but also impact the sensitivity to finite size effects. The latter is due to the fact that  the geometric spread of the accumulated induced charge is impacted by the chemistry of the membrane surface and the pore entrance. Comparing the passage times and free energy barriers computed for both systems reveals that the ion transport appears to be more restricted for the pore chemistry considered here, resulting in an increase in $\tau_c$ from $2.64\pm0.16~\mu$s in Ref.~\citenum{MalmirMatter2020} to $13.6\pm1.7~\mu$s here. Applying the ideal conductor model, however, leads to $\tau_{c,\infty}$ values that are almost statistically indistinguishable ($126\pm36$~ns for Ref.~\citenum{MalmirMatter2020} vs.~$322\pm117$~ns for the current work). This is a vivid demonstration of the fact that the magnitude of the finite size correction to the generalized free energy profile can be system-dependent due to the sensitivity of $z_c$ to the membrane and pore chemistry. 

\begin{figure*}
	\centering
	\includegraphics[width=.999\textwidth]{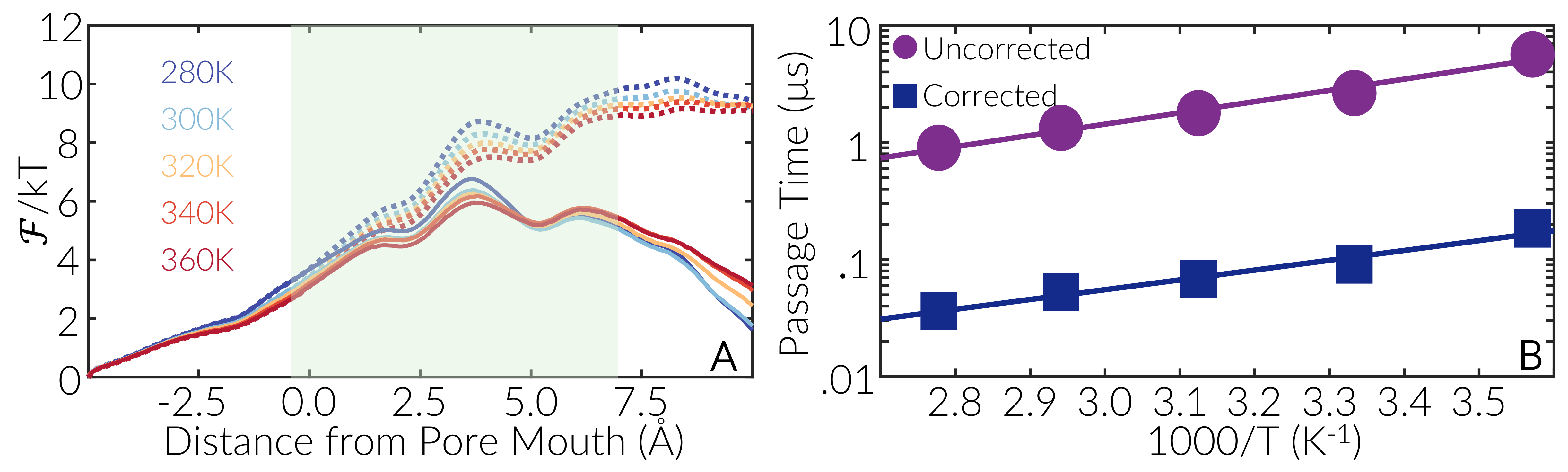}
	\caption{\label{fig:Correct-Hessam} (A) $\mathcal{F}_f(\lambda)$ (dashed) and $\mathcal{F}_\infty(\lambda)$ (solid) profiles computed for the membrane considered in Ref.~\citenum{MalmirMatter2020}. (B) Arrhenius plots for the original (dark purple) and corrected (dark blue) $\tau_c$ values. } 
\end{figure*}

Figure~\ref{fig:Correct-Hessam}A depicts the original and corrected free energy profiles for the system considered in Ref.~\citenum{MalmirMatter2020}. Unlike the pore chemistry considered in this work, the dominant free energy peak observed in finite-size simulations does not fully disappear in the thermodynamic limit, and only becomes less dominant in comparison to the peak at the middle of the pore. It is  also noteworthy  that the $\tau_{c,\infty}$'s estimated at different temperatures exhibit an Arrhenius dependence on $T$ with an activation barrier of $\Delta E_{c,\infty}=16.0\pm4.7$~kJ/mol. Note that $\Delta E_{c,\infty}$ is slightly smaller than $\Delta E_{c,f}=18.4\pm4.4$~kJ/mol (Fig.~\ref{fig:Correct-Hessam}B), but is still larger than the activation barrier for water transport.

\section{Discussions \& Conclusions}

\noindent
This work constitutes the first systematic investigation of how the size of the simulation box will impact the timescales of water and ion transport through semipermeable nanoporous membranes under non-equilibrium and driven conditions.
Similar to our earlier work,\cite{MalmirMatter2020} we use a thin graphite-based membrane with a sub-nm cylindrical pore as a test case and vary the lateral dimensions of the membrane while keeping the pore size and chemistry and the compartment concentrations fixed. By using \textsc{Nemd} and j\textsc{Ffs}, we compute the mean first passage times for water molecules and chloride ions  as a function of the system size. While water fluxes are virtually insensitive to system size, ionic transport timescales vary by almost six orders of magnitude and decrease as the system size increases. Our examination of generalized free energy barriers (estimated from the \textsc{Ffs}-\textsc{Mfpt} method)  points not only to a wide $13kT$ spread in the free energy barrier, but also to a qualitative change in the perceived mechanism of transport. For smaller systems, $\mathcal{F}(\lambda)$ exhibits two major peaks, with the higher one located right outside the pore exit. As the system size increases, however, the second peak weakens and eventually disappears. As a result, the locus of the main barrier moves discontinuously from outside the pore exit to the middle of the pore. 

By carefully examining the force exerted on the leading ion, we attribute these intriguing observations to the differential distribution on the membrane surface of the surplus charge induced within the feed compartment as a result of the leading ion's entrance into the pore. To better quantify the observed finite size artifacts, we develop a theoretical model assuming that the feed compartment is an ideal conductor. Despite the stark quantitative and qualitative differences between the generalized free energy profiles computed for different system sizes, subtracting the correction term predicted by our theoretical model yields free energy profiles that collapse onto one another and become statistically indistinguishable. Our physics-based approach therefore constitutes  a blueprint for rigorously removing finite-size artifacts from the fluxes and free energy profiles obtained from computational studies of hindered ion transport.

It must be emphasized that the scaling of activation barriers and ionic passage times with system size does not only depend on the dimensions of the simulation box, but is also likely to be impacted by other features, such as the size, arrangement and chemistry of the pores. When it comes to our ideal conductor theory, these structural features impact the thickness and location of what constitutes the surface of the ideal conductor and henceforth the $z_c$ parameter. Therefore, comparing the passage times and free energy barriers of two or more finite systems with identical box dimensions but different pore chemistries and arrangements without applying finite size corrections can be perilous and might lead to incorrect and misleading conclusions. This can be vividly observed in comparing the two otherwise identical pores that only have slightly different passivation patterns, namely the pores considered in our earlier work\cite{MalmirMatter2020} and those considered here.  While ion transport appears to be almost an order of magnitude faster through the former, the two membranes tend to have almost identical free energy barriers and passage times in the thermodynamic limit. This is therefore an important point that needs to be taken into account in studies that involve probing a design parameter space.

Apart from their importance in making molecular simulations of ionic transport through membranes more accurate, our findings might shed some light onto apparent discrepancies between experimental and computational studies of ion transport,~e.g.,~in water desalination. While most computational studies report salt rejection rates of 100\%,\cite{HeiranianNatComm2015, CohenNanoLett2016, YanNanoscale2017, ZhangPCCP2020} experimental studies generally yield  smaller rejection rates.\cite{ChildressEnvSciTechnol2000, RaziEnvSciTech2018, PangJColloidInterSci2018, LiDesalination2019, YaoNatSust2021} Part of this disagreement is due to defects in real membranes,\cite{LeeJMembrSci2011} but our work suggests that the finite sizes of the simulation boxes considered in simulation studies might also lead to massive overestimation of ionic passage times and salt rejection rates. Considering the linear scaling of  $\Delta\mathcal{F}_{\text{corr}}$ with $z$ as $z\rightarrow\infty$, such discrepancies are expected to be larger for thicker membranes. 

While estimating the precise correction term to be subtracted from $\mathcal{F}(\lambda)$ requires applying the full ideal conductor model, a conservative upper bound can still be obtained from the dimensions of the membrane and its dielectric constant. More specifically, assuming the worst case scenario in which the transition state is located right at the membrane exit (i.e.,~on the filtrate side), and that the excess charge is almost uniformly distributed at the membrane surface, $\Delta\mathcal{F}^*_{\text{corr}}$, the correction term at the top of the activation barrier, will bounded from above by
\begin{eqnarray}
\frac{\Delta\mathcal{F}^*_{\text{corr}}}{kT} &\le& \underbrace{\frac{Z^2{\text{\Large \calligra e}\,}^2}{2\epsilon_0 kT}}_{\alpha} \underbrace{\frac{l_m}{L_xL_y\langle \epsilon_r\rangle}}_{\kappa} ,
\label{eq:fe-corr-upper-bound}
\end{eqnarray}
where $Z$ is the valence of the traversing ion and $l_m$ is the thickness of the membrane. Note that  $\alpha$ only depends on temperature, while $\kappa$ is a function of the  dimensions of the membrane as well as the normal component of its  dielectric constant tensor. Therefore,  Eq.~(\ref{eq:fe-corr-upper-bound}) can be used as a heuristic to \emph{a priori} choose $L_x$ and $L_y$ so that $\Delta{F}_{\text{corr}}^*$ is smaller than an acceptable threshold,~e.g.,~a few $kT$'s. 

The framework proposed here can, in principle, be  applied to the problem of sodium transport,~i.e.,~to the $F_{0,1}\rightarrow F_{1,1}$ transition in which the first sodium traverses the pore after a chloride has already reached the filtrate. Qualitatively, one would expect the sodium passage time to increase dramatically as the system size increases, since  the leading sodium will be under an excess electrostatic ``pull-forward'' force even at $F_{0,1}$,~i.e.,~while still inside the feed compartment. Therefore, the $F_{0,1}$ basin will be further destabilized in comparison to the top of the barrier. There are, however, several outstanding technical obstacles to quantifying this effect. First of all, the destabilization of the $F_{0,1}$ implies that a curvature correction needs to be applied to resolve the mismatch between the curved directed distance from the pore entrance and the perpendicular distance from the membrane surface.  The more daunting challenge, however, is the need to estimate $\epsilon_r(z)$, the $z$-dependence normal component of the dielectric constant tensor at the immediate vicinity of the membrane surface. As has been demonstrated before, dielectric constants change drastically upon confinement. In the case of  water, for instance, the dielectric constant can change by over an order of magnitude with respect to the bulk.\cite{SchlaichPRL2016, FumagalliScience2018} Moreover, as discussed elsewhere, it is not straightforward to calculate anisotropic and position-dependent dielectric constants, as the existing methodologies can also exhibit strong finite size effects even when dealing with an insulating medium such as pristine water.\cite{OlivieriJPCL2021} Exploring these technical challenges and developing robust algorithms for estimating position-dependent dielectric constants is necessary for correcting the passage times and generalized free energy profiles for the $F_{0,1}\rightarrow F_{1,1}$ transition. Addressing these challenges is beyond the scope of this work and they will be topics of follow-up studies.

One of the key underlying assumptions of the ideal conductor model is that charge carriers within the conductor are infinitesimally small. This condition, however, is not satisfied in molecular simulation wherein a finite number of charge carriers are present. This non-ideality appears to have no noticeable impact on the performance of the ideal conductor model here. This might, however, no longer be the case in situations in which the excess or deficit charge is carried by a small number of charge carriers. For instance, such non-ideality can have profound impacts on correcting for finite size effects in the $F_{0,1}\rightarrow F_{1,1}$ transition wherein the deficit charge in the filtrate is only carried by a single ion. Further follow-up studies are needed for determining the conditions under which the small number of charge carriers might impact the predictive power of the ideal conductor model.

Another key assumption of the ideal conductor model is that charge carriers can reorganize instantaneously in response to electrostatic perturbations. In real systems, however, there is always a characteristic timescale for the diffusion of charge carriers, resulting in a gap between the stimulus (the motion of the leading ion) and the response (charge rearrangement at the membrane surface). A reasonable proxy for such a gap is the cage escape time, which can be evaluated as $\tau_{\text{cage}}=l_r^2/D$ where $l_r$ and $D$ are the locus of the first valley of the radial distribution function and the diffusivity, respectively. We use the Einstein relationship to compute the ionic diffusivities from mean-squared displacements (SI Fig.~S2). Using the computed diffusivities ($D_{\text{chloride}}=1.40\times10^{-9}~\text{m}^2\cdot\text{s}^{-1}$ and $D_{\text{sodium}}=1.98\times10^{-9}~\text{m}^2\cdot\text{s}^{-1}$) and the loci of the first valleys of RDF obtained in Ref.~\citenum{MalmirMatter2020} ($l_{r,\text{sodium}}=0.325~\text{nm}$ and $l_{r,\text{chloride}}=0.375~\text{nm}$), we estimate the cage escape times to be $75$ and $71$~ps for sodium and chloride, respectively. These are both comparable in magnitude to the typical length of a trial \textsc{Ffs} trajectory (a few tens of picoseconds). This implies that the lag between stimulus and the response is negligible here. Such a lag, however, might become significant when the feed is viscous or when charge carriers are bulkier and have considerably smaller diffusivities. The latter can, for instance, happen in solutions of charged polymers or biomolecules. Another situation that might make such lags important is when a system experiences oscillatory electric fields. Determining a charge carrier diffusivity below which such lags becomes important is beyond the scope of this work and will be the topic of future studies.

It must also be noted that our model is a powerful framework for studying the transport of ions and charge macromolecules through biological membranes and channel proteins. It has indeed been demonstrated that ion transport  through channel proteins will be significantly slower when non-polarizable force-fields are utilized.\cite{NgoJCTC2021} While such apparent slowdown could be partly caused by the limitations of non-polarizable force-fields, they might also arise from the types of finite size effects discussed in this work. Therefore, even though polarizable force-fields might be less susceptible to such artifacts, equally accurate estimates of passage times might still be attainable from non-polarizable models if finite size effects are properly corrected.

\section*{Supporting Information Available}
\noindent
Supporting information contains two figures.\\

\section*{ACKNOWLEDGMENTS}
\noindent
A.H.-A. gratefully acknowledges the support of the National Science Foundation Grants CBET-1751971 (CAREER Award) and CBET-2024473. B.S. acknowledges the support of the Goodyear Tire \& Rubber Fellowship.  We thank P. G. Debenedetti, M. Elimelech, V. Batista, D. Laage and R. Coifmann for useful discussions. These calculations were performed on the Yale Center for Research Computing. This work used the Extreme Science and Engineering Discovery Environment (XSEDE), which is supported by National Science Foundation grant no. ACI-1548562~\cite{TownsCompSciEng2014}.

\appendix

\section{Derivation of Eq.~(\ref{eq:MFPT})}
\label{secction:jumpy-correction}

\noindent
In order to prove Eq.~(\ref{eq:MFPT}), we use notations similar to the ones introduced in our earlier paper.\cite{HajiAkbariJChemPhys2018} 
In particular, for a discrete-time \textsc{Nemd} trajectory $X\equiv(x_0,x_1,x_2,\cdots)$, we reuse the following random variables and functions already introduced in Ref.~\citenum{HajiAkbariJChemPhys2018}:
\begin{eqnarray}
L[X] &=& \min_{q>0}\{ x_q\not\in A\}\\
T_B[X] &=& \min_{q\ge L[X]} \{x_q\in A\cup B\}\\
W_B[X] &=& \Theta_B\left(
x_{T_B[X]}
\right)\\
T_i[X] &=& \min_{q\ge L[X]} \left\{
x_q\not\in\cup_{j=0}^i\mathcal{C}_{j-1}
\right\}\\
W_i[X] &=& \theta_i\left(
x_{T_i[X]}
\right)
\\
U_{i,j}[X] &=& \left\{
\begin{array}{ll}
\theta_i\left(x_{T_i[X]}\right)\theta_j\left(x_{T_{i+1}[X]}\right) & i\ge 0\\
\phi_0\left(x_{L[X]}\right)\theta_j\left(x_{T_0[X]}\right) & i=-1
\end{array}
\right.\\
\phi_i(x) &=& \sum_{j=0}^i\theta_{j-1}(x)
\end{eqnarray}
Here, $\mathcal{C}_i=\{x_\mathcal{Q}: \lambda_i\le \lambda(x) < \lambda_{i+1}\}$, while $\theta_B(x)$ and $\theta_i(x)$ are the indicator functions for $B$ and $\mathcal{C}_i$, respectively. Moreover, for any $\lambda_k<\lambda\le\lambda_{k+1}$, we define the following random variables:
\begin{eqnarray}
T_{\lambda}[X] &=& \min_{q\ge L[X]} \{\lambda(x_q)\ge\lambda,~\text{or}~x_q\in A \}
\\
W_\lambda[X] &=& \left\{
\begin{array}{ll}
1 & \lambda\left(x_{T_\lambda[X]}\right)\ge\lambda \\
0 & x_{T_\lambda[X]} \in A
\end{array}
\right.
\\
U_{k-1,\lambda}[X] &=& \left\{
\begin{array}{ll}
1 & \lambda\left(
x_{T_k[X]}
\right)\ge\lambda\\
0 & \text{otherwise}
\end{array}
\right. 
\end{eqnarray}
For simplicity, we assume that \textsc{Ffs} milestones are placed so that the simplified variant of j\textsc{Ffs} introduced in Ref.~\citenum{HajiAkbariJChemPhys2018} can be employed,~i.e.,~that no multi-milestone jumps are observed. This will imply that $U_{i,j}$ will always be zero for $j>i+1$. This assumption considerably simplifies the procedure for computing $\langle T_\lambda\rangle$ and $\langle W_\lambda\rangle$, which are both needed for estimating  the $\tau(\lambda;A)$ given in Eq.~(\ref{eq:MFPT}). In particular, $\langle T_\lambda\rangle$ can be estimated as:
\begin{eqnarray}
\langle T_\lambda\rangle_{\mathcal{E}_A} &=& \langle T_k\rangle_{\mathcal{E}_A} + \notag\\
&&   \langle T_\lambda - T_k| U_{k-1,\lambda}=1\rangle_{\mathcal{E}_A} P\left(
U_{k-1,\lambda}=1
\right)+\notag\\
&& \langle T_\lambda - T_k| U_{k-1,\lambda}=0\rangle_{\mathcal{E}_A} P\left(
U_{k-1,\lambda}=0
\right)\notag\\
&\overset{\text{(a)}}{=}& \langle T_k\rangle_{\mathcal{E}_A}  
\notag\\
&& +\langle T_\lambda-T_k|W_\lambda=1,U_{k-1,\lambda}=0, \textbf{U}_k=\pmb1\rangle_{\mathcal{E}_A}\times\notag\\
&&  P\left(
W_\lambda=1, U_{k-1,\lambda}=0, \textbf{U}_k=\pmb1
\right) \notag\\
&& +  \langle T_\lambda-T_k|W_\lambda=0,U_{k-1,\lambda}=0, \textbf{U}_k=\pmb1\rangle_{\mathcal{E}_A}\times\notag\\
&&  P\left(
W_\lambda=0, U_{k-1,\lambda}=0, \textbf{U}_k=\pmb1
\right) 
\notag\\
&=& \langle T_k\rangle_{\mathcal{E}_A} + P(\textbf{U}_k=\pmb1)\times\notag\\
&&\left(1-F_{k}^{\lambda^+}\right)\left[
L_\lambda^{(s)}\frac{S_\lambda}{M_k^{\lambda^+}}
+L_{\lambda}^{(f)}\frac{M_k^{\lambda^+}-S_\lambda}{M_k^{\lambda^+}}
\right]\notag\\
&& \label{eq:bar-T-lambda}
\end{eqnarray}
Here $\textbf{U}_k\equiv = (U_{-1,0},U_{0,1}, U_{1,2}, \cdots, U_{k-1,k})$, $\pmb1$ is a $(k+1)$-dimensional vector of ones, $\mathcal{E}_A$ is the space of all trajectories initiated from $A$, $F_k^{\lambda^+}$ is the fraction of crossing events at $\lambda_k$ that land at a value $\ge\lambda$, $M_{k}^{\lambda^+}$ is the number of trial trajectories initiated from the configurations landed before $\lambda$ and $S_\lambda$ is the number of those that reach $\lambda$. $L_\lambda^{(s)}$ and $L_\lambda^{(f)}$ correspond to the average length of successful and failing trajectories (computed for the $M_k^{\lambda^+}$ trajectories launched from configurations prior to $\lambda$), respectively wherein success corresponds to reaching $\lambda$.  Note that (a) follows from the fact that $T_k[X]=T_{\lambda}[X]$ when $U_{k-1,\lambda}[X]=1$ or $\textbf{U}_k\neq\pmb1$. Similarly, $P(W_\lambda=1)$ can be estimated as,
\begin{eqnarray}
P(W_\lambda=1)  &=& P(W_\lambda=1, \textbf{U}_k=\pmb1)  \notag\\
&=& \sum_{b=0}^1 P(W_\lambda=1, \textbf{U}_k=\pmb1, U_{k-1,\lambda} = b) 
\notag\\
&=& \sum_{b=0}^1 P(W_\lambda=1|U_{k-1,\lambda}=b, \textbf{U}_k=\pmb1)\times\notag\\
&& P(U_{k-1,\lambda}=b|\textbf{U}_k=\pmb1) P(\textbf{U}_k=\pmb1)
\notag\\
&=& \left[
\frac{S_\lambda}{M_k^{\lambda^+}}\left(1-F_k^{\lambda^+}\right) + 1\cdot F_k^{\lambda^+}
\right]P(\textbf{U}_k=\pmb1)\notag\\
&&\label{eq:bar-P-lambda}
\end{eqnarray}
Eq.~(\ref{eq:MFPT}) can then be derived from Eqs.~(\ref{eq:bar-T-lambda}) and (\ref{eq:bar-P-lambda}) by noting that $\tau(\lambda;A)={\langle T_\lambda\rangle_{\mathcal{E}_A}}/{\langle W_\lambda\rangle_{\mathcal{E}_A}}$.

%

\section{Derivation of Eqs.~(\ref{eq:main-superposition-reciprocal}) and~(\ref{eq:fe-corr})}
\label{section:correction}

\noindent
Suppose that a monovalent anion is removed from an ideal conducting slab that is infinitely large in the $x$ and $y$ dimensions and that has its top surface at $z=0$. Assuming that the anion is located at $(0,0,h)$, the charge density induced by its removal can be readily estimated from the method of images and is given by Eq.~(\ref{eq:rho-inf}).
Now suppose that the anion is removed from a finite ideal conductor (with lateral dimensions $L_x$ and $L_y$) that is also periodic along the $x$ and $y$ dimensions. Under such a scenario, all the periodic images of that ion will also leave the conductor simultaneously, and the resulting induced charge density can be estimated from the superposition principle:
\begin{eqnarray}
\rho_f(x,y) &=& \frac{\text{\calligra\Large e}\,h}{2\pi} \sum_{k_x,k_y=-\infty}^{+\infty} \frac{1}{\left[
(x-k_xL_x)^2+(y-k_yL_y)^2+h^2
\right]^{\frac32}} \notag\\&=& \sum_{k_x,k_y\in\mathbb{Z}} \rho_{\infty}(x-k_xL_x, y-k_yL_y)
\label{eq:superposition-real}
\end{eqnarray}
It must be noted that $\rho_f(x,y)$ is periodic along the $x$ and $y$ directions and can thus be expressed as the following Fourier series:
\begin{eqnarray}
\rho_f(x,y) &=& \frac{1}{L_xL_y} \sum_{k_x,k_y} e^{i(q_xx+q_yy)}\times\notag\\&&\underbrace{\int_0^{L_x} \int_0^{L_y} e^{-i(q_x\xi+q_y\eta)}\rho_f(\xi,\eta) d\xi d\eta}_{a_{k_x,k_y}}\notag\\&&
\label{eq:Fourier-series}
\end{eqnarray}
Note that $\textbf{q}=(2\pi k_x/L_x, 2\pi k_y/L_y)\in\mathbb{R}^2$ is the wavevector associated with $(k_x,k_y)\in\mathbb{Z}^2$. The goal is to use the symmetries of $\rho_\infty(x,y)$ and $\rho_f(x,y)$ to estimate $a_{k_x,k_y}$ and to express $\rho_f(x,y)$ in the reciprocal space. By combining Eqs.~(\ref{eq:superposition-real}) and (\ref{eq:Fourier-series}), it can be demonstrated that:
\begin{eqnarray}
a_{k_x,k_y} &=& \sum_{j_x,j_y\in\mathbb{Z}}\int_0^{L_x} \int_0^{L_y} e^{-i(q_x\xi+q_y\eta)}\times\notag\\
&&\rho_\infty(\xi-j_xL_x,\eta-j_yL_y)\,d\xi d\eta\notag\\
&=& \sum_{j_x,j_y\in\mathbb{Z}} \int_{-j_xL_x}^{(1-j_x)L_x}\int_{-j_yL_y}^{(1-j_y)L_y} e^{-i(q_x\overline{\xi}+q_y\overline{\eta})}\times\notag\\
&&\rho_\infty(\overline{\xi},\overline{\eta}) d\overline{\xi} d\overline{\eta}\notag\\
&=& \int_{-\infty}^{+\infty}\int_{-\infty}^{+\infty}e^{-i(q_x\overline{\xi}+q_y\overline{\eta})}\rho_\infty(\overline{\xi},\overline{\eta}) d\overline{\xi} d\overline{\eta} \notag\\
&=& \widehat{\rho_\infty}\left(\frac{\textbf{q}}{2\pi}\right)
\label{eq:Fourier-coeffs}
\end{eqnarray}
Here, $\widehat{\rho_\infty}(\cdot)$ is the Fourier transform of $\rho_f(\cdot)$, which can be readily calculated by noting that $\rho_\infty(x,y)$ is a radial function,~i.e.,~that $\rho_\infty(x,y)=f\left(\sqrt{x^2+y^2}\right)$. Therefore, according to the  scaling theorem for Fourier transforms,\cite{FourierAnalSteinShekarchi2011} $\widehat{\rho_\infty}\left({q_x}/{2\pi},{q_y}/{2\pi}\right)$ will also be a radial function in the reciprocal space,~i.e.,~it will only depend on  $|\textbf{q}|=\sqrt{q_x^2+q_y^2}$. Therefore, for any $\textbf{q}=(q_x,q_y)$, it suffices to estimate $\widehat{\rho_\infty}(\textbf{q})$ for $(|\textbf{q}|,0)$.  $\widehat{\rho_\infty}(\textbf{q})$ will thus be given by,
\begin{eqnarray}
\widehat{\rho_\infty}\left(\frac{\textbf{q}}{2\pi}\right) &=& \frac{\text{\calligra\Large e}\,h}{2\pi}\int_{-\infty}^{+\infty} \int_{-\infty}^{+\infty} \frac{e^{-i|\textbf{q}|\xi}}{\left[\xi^2+\eta^2+h^2\right]^{\frac32}} d\eta d\xi\notag \\
&=& \frac{\text{\calligra\Large e}\,h}{\pi}{\int_{-\infty}^{+\infty} \frac{e^{-i|\textbf{q}|\xi}}{\xi^2+h^2}d\xi} \overset{\text{(a)}}{=}  \text{\calligra\Large e}\,e^{-|\textbf{q}|h} 
\end{eqnarray}
Note that (a) can be proven by conducting a contour integral and using the residue theorem. $\rho_f(x,y)$ can therefore be expressed as
\begin{eqnarray}
\rho_f(x,y) &=& \frac{\text{\calligra\Large e}}{L_xL_y}\sum_{k_x,k_y\in\mathbb{Z}} e^{i(q_xx+q_yy)}e^{-|\textbf{q}|h} \notag\\
&\overset{\text{(a)}}{=}& \frac{\text{\calligra\Large e}}{L_xL_y}\sum_{k_x,k_y\in\mathbb{Z}} \cos (q_xx+q_yy) e^{-|\textbf{q}|h}\label{eq:rhof-raw}
\end{eqnarray}
(a) follows from the fact that $\rho_f(x,y)$ is real-valued and thus $\rho_f(x,y)=\operatorname{Re}\rho_f(x,y)$. Eq.~(\ref{eq:main-superposition-reciprocal}) can thus be obtained by algebraically rearranging Eq.~(\ref{eq:rhof-raw}) by noting that  the corresponding sum is invariant under inversion. 
The next step is to obtain the $z$ component of the electric field induced by the charge density given by Eq.~(\ref{eq:main-superposition-reciprocal}):
\begin{eqnarray}
\widetilde{E}_{f,z}^{h}(0,0,z) &=& \frac{2\epsilon_0\epsilon_r(z)L_xL_yE_{f,z}^h(0,0,z)}{\text{\calligra\Large e}} \notag\\
&=& \frac{h}{2\pi}\int_{-\infty}^{+\infty}\int_{-\infty}^{+\infty} \frac{z\widetilde{\rho}_f(x,y)dxdy}{\left[x^2+y^2+z^2\right]^{\frac32}} \notag\\
&\overset{\text{(a)}}{=}& 1 + 2\sum_{k=1}^{+\infty} \left[
e^{-\frac{2\pi(z+h)k}{L_x}} + e^{-\frac{2\pi(z+h)k}{L_y}}
\right] \notag\\
&& +4\sum_{k_x,k_y=1}^{+\infty} e^{-2\pi(z+h)\sqrt{\frac{k_x^2}{L_x^2}+\frac{k_y^2}{L_y^2}}} 
\label{eq:dimensionless-field}
\end{eqnarray}
Here, $\epsilon_0$ and $\epsilon_r(z)$ are the vacuum permittivity and the local dielectric constant of the material, respectively. Note that (a) follows from using the following integration identity that can also be proven using the residue theorem:
\begin{eqnarray}
\int_{-\infty}^{+\infty} \int_{-\infty}^{+\infty} \frac{\cos\frac{2\pi kx}{L}}{\left[x^2+y^2+a^2\right]^{\frac32}}dxdy &=& \frac{2\pi}{a} e^{-\frac{2\pi ka}{L}}
\end{eqnarray}
The next step is to integrate Eq.~(\ref{eq:dimensionless-field}) to obtain the excess change in the electrostatic potential due to the presence of $\widetilde{\rho}_f(x,y)$. More precisely, the excess electrostatic potential, $\Phi^{\text{ex}}(h)$ is given by:
\begin{eqnarray}
\widetilde{\Phi}^{\text{ex}}(h) - \widetilde{\Phi}^{\text{ex}}(h_0)&=& \frac{2\epsilon_0 L_xL_y\left[\Phi^{\text{ex}}(h)-\Phi^{\text{ex}}(h_0)\right]}{\text{\calligra\Large e}\,h} \notag\\
&=& \bigintsss_{h_0}^h \frac{\left[\widetilde{E}_{f,z}^h(0,0,z)-\widetilde{E}_{\infty,z}^h(0,0,z)\right]\,dz}{h\epsilon_r(z)} 
\notag\\ &&\label{eq:potential-general}
\end{eqnarray}
Here, $h_0$ is the point at which the anion fully detaches from the conductor and the charge imbalance (resulting in $\rho_f(x,y)$) is established. In this work, we use $h_0=0$, which corresponds to when the leading ion alongside its first hydration shell fully leaves the feed. Assuming that $\epsilon_r(z)\equiv1$, which is a reasonable approximation for the interior of the membrane considered in this work, Eq.~(\ref{eq:potential-general}) can be simplified as:
\begin{eqnarray}
\widetilde{\Phi}^{\text{ex}}(h)&=&  \widetilde{\Phi}^{\text{ex}}(h_0) + \frac{h_0}{h}-1  \notag\\
&& +\frac{L_x}{\pi h} \sum_{k=1}^{+\infty} \frac{e^{-4\pi kh/L_x}- e^{-2\pi k(h+h_0)/L_x}}{k} \notag\\
&& +\frac{L_y}{\pi h} \sum_{k=1}^{+\infty} \frac{e^{-4\pi kh/L_y}- e^{-2\pi k(h+h_0)/L_y}}{k} \notag\\
&& + \frac{2}{\pi h} \sum_{k_x,k_y=1}^{+\infty} \frac{1}{\sqrt{(k_x/L_x)^2+(k_y/L_y)^2}}\times \notag\\
&& \Bigg[e^{-4\pi h\sqrt{(k_x/L_x)^2+(k_y/L_y)^2}}\notag\\
&& -e^{-2\pi(h+h_0)\sqrt{(k_x/L_x)^2+(k_y/L_y)^2}}\Bigg]\notag\\
&& +\frac{L_xL_y}{2\pi h}\left[\frac{1}{h_0+h}-\frac1{2h}\right]
\end{eqnarray}
The excess electrostatic potential can be used to estimate the correction in free energy given in Eq.~(\ref{eq:fe-corr}).

\bibliographystyle{apsrev}
\bibliography{References}

\clearpage
\setcounter{figure}{0}
\setcounter{table}{0}
\renewcommand{\thefigure}{S\arabic{figure}}
\renewcommand{\thetable}{S\arabic{table}}

\section*{SUPPLEMENTARY INFORMATION}

\allowdisplaybreaks

\begin{figure*}
	\centering
	\includegraphics[width=.37\textwidth]{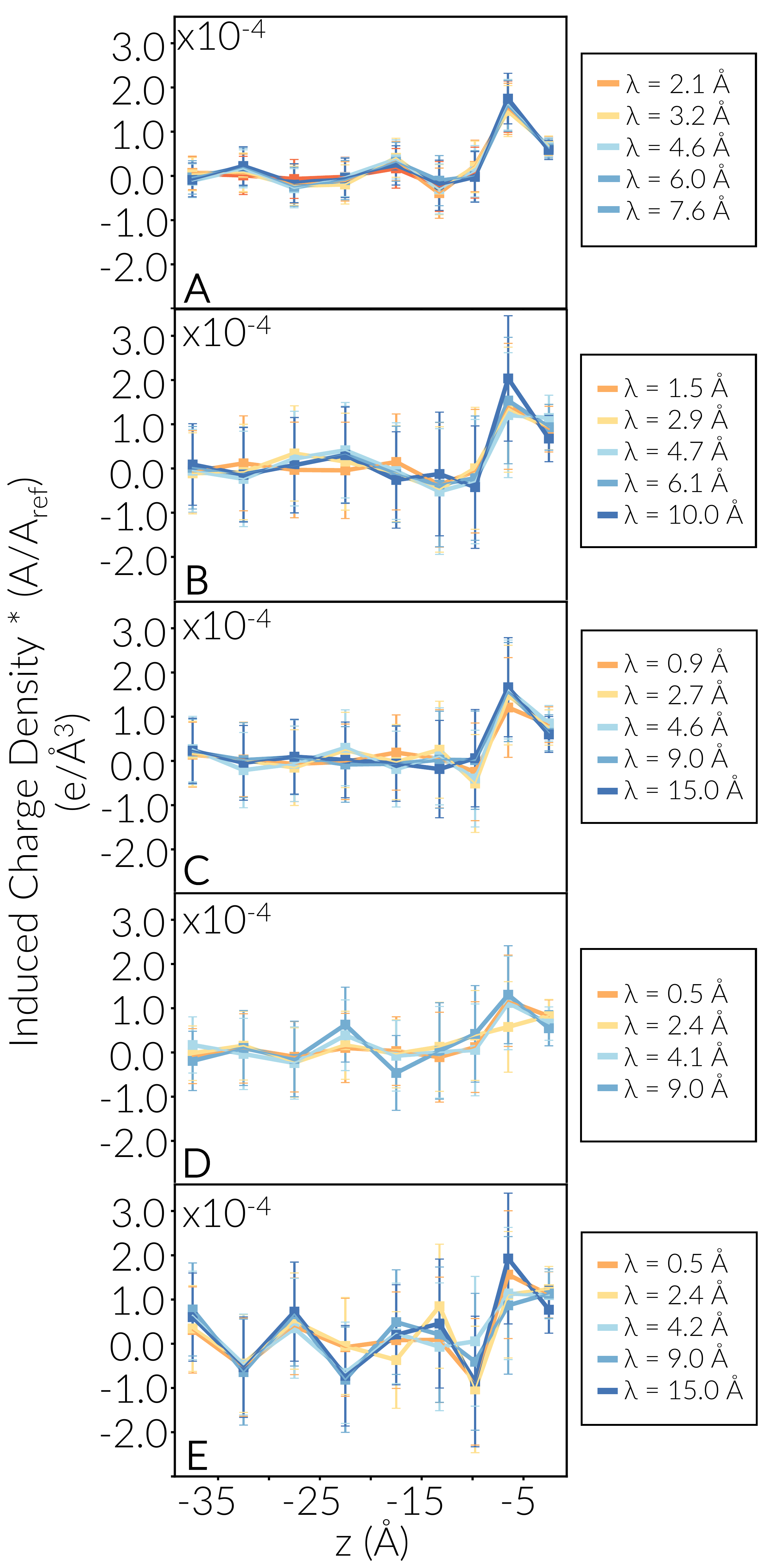}
	\caption{The induced charge density profiles within the feed compartment as a function of distance from the membrane surface for different \textsc{Ffs} milestones in the system with membrane cross-sectional area of: (A) $12.54~\text{nm}^2$, (B) $26.33~\text{nm}^2$, (C) $32.60~\text{nm}^2$, (D) $51.82~\text{nm}^2$ and (E) $102.81~\text{nm}^2$. The charge densities have all been rescaled by the membrane surface area $A/A_{\text{ref}}$ to facilitate comparisons  across different system sizes, where $A_{\text{ref}}$ is the membrane cross-sectional area of the smallest system.}
\end{figure*}

\begin{figure*}
	\centering
	\includegraphics[width=.7\textwidth]{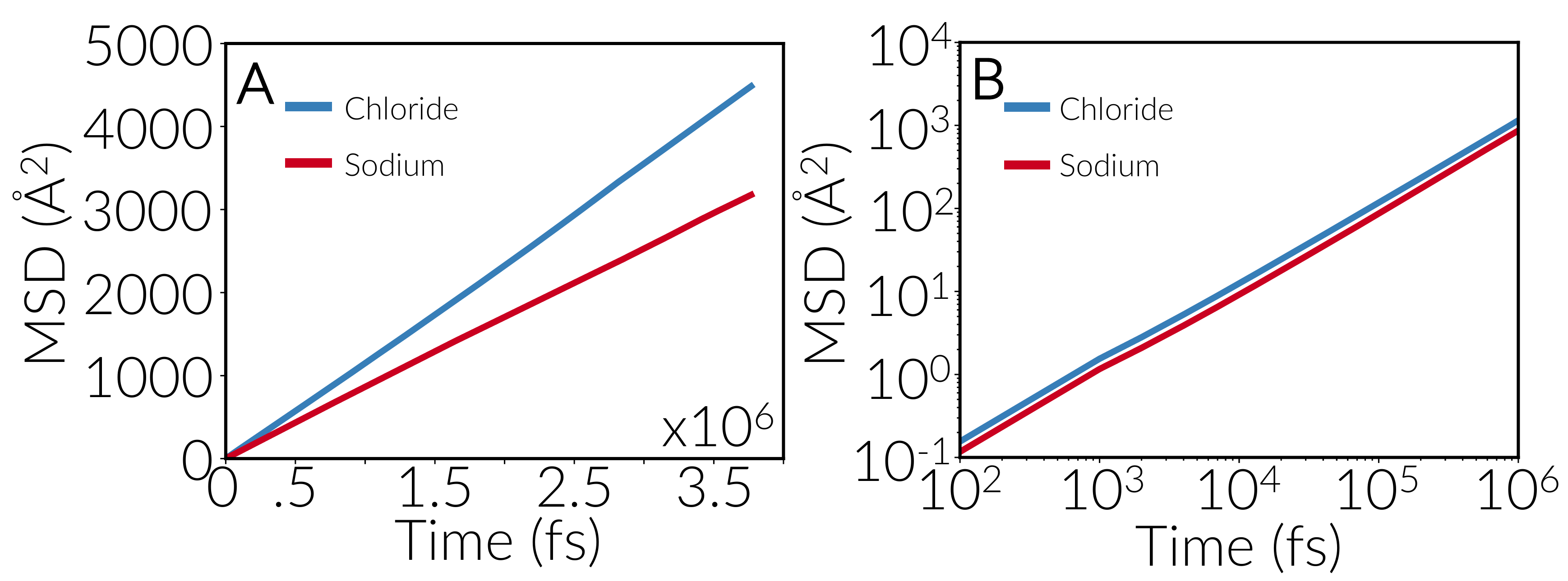}
	\caption{Mean-squared displacement of the sodium and chloride ions computed from \textsc{Nvt} simulations of bulk NaCl solution at 300~K depicted in (A) linear and (B) log-log scales. The diffusivities of the sodium and chloride ions-- as computed from the Einstein relationship-- are $1.40\times10^{-9}~\text{m}^2\cdot\text{s}^{-1}$ and $1.98\times10^{-9}~\text{m}^2\cdot\text{s}^{-1}$, respectively. }
\end{figure*}

\end{document}